\documentclass[journal]{IEEEtran}
\usepackage{algorithm}
\usepackage{algorithmic}
\usepackage{ifpdf}
\usepackage{cite}
\ifCLASSINFOpdf
  \usepackage[pdftex]{graphicx}
\else
  \usepackage[dvips]{graphicx}
\fi
\usepackage{graphicx}

\usepackage{amsmath}
\usepackage{ amssymb }
\usepackage{algorithmic}
\usepackage{array}
\ifCLASSOPTIONcompsoc
  \usepackage[caption=false,font=normalsize,labelfont=sf,textfont=sf]{subfig}
\else
  \usepackage[caption=false,font=footnotesize]{subfig}
\fi

\ifCLASSOPTIONcaptionsoff
  \usepackage[nomarkers]{endfloat}
 \let\MYoriglatexcaption\caption
 \renewcommand{\caption}[2][\relax]{\MYoriglatexcaption[#2]{#2}}
\fi
\usepackage{url}
\usepackage{ dsfont }
\usepackage{color}

\newcommand{\blue}[1]{\textcolor[rgb]{0,0,1}{#1}}

\begin{document}

\title{A Data-driven Storage Control Framework \\for Dynamic Pricing}

\author{Jiaman Wu, Zhiqi Wang, Chenye Wu, Kui Wang, \emph{and} Yang Yu
\thanks{The authors are with the Institute for Interdisciplinary Information Sciences (IIIS), Tsinghua University, Beijing, China, 100084. C. Wu is the correspondence author. Email: chenyewu@tsinghua.edu.cn.}
}

\maketitle

\begin{abstract}
Dynamic pricing is both an opportunity and a challenge to the demand side. It is an opportunity as it better reflects the real time market conditions and hence enables an active demand side. However, demand's active participation does not necessarily lead to benefits. The challenge conventionally comes from the limited flexible resources and limited intelligent devices in demand side. The decreasing cost of storage system and the widely deployed smart meters inspire us to design a data-driven storage control framework for dynamic prices. We first establish a stylized model by assuming the knowledge and structure of dynamic price distributions, and design the optimal storage control policy. Based on Gaussian Mixture Model, we propose a practical data-driven control framework, which helps relax the assumptions in the stylized model. Numerical studies illustrate the remarkable performance of the proposed data-driven framework.
\end{abstract}

\vspace{0.1cm}

\IEEEpeerreviewmaketitle

\IEEEpeerreviewmaketitle
\section{Introduction}
\label{sec::1intro}
\IEEEPARstart{I}n terms of economic effectiveness, dynamic pricing outperforms its rivals since it accurately represents the value of energy in real time. More importantly, it provides proper incentives to the demand side to achieve a cost effective power system \cite{borenstein2002dynamic}. 

Nonetheless, dynamic pricing does not necessarily benefit every end user. After implementing dynamic price in Alberta, most end users, who opted in variable electricity rates, witnessed dramatic increase in their electricity bills during winter over the past few years \cite{energyrate}.

On the one hand, this highlights that dynamic price does reflect the true value of energy in real time. On the other hand, such price signal, due to limited intelligence in the demand side, only poses risks to most end users. To better utilize the advantage of dynamic pricing, low cost storage systems and smart meters with intelligent control capabilities are of vital importance. Together with these two necessary components, dynamic pricing can enable an active demand side, and can thus achieve the system-wide economic effectiveness.

\begin{figure}[t]
\centering
\includegraphics[width=2.9in]{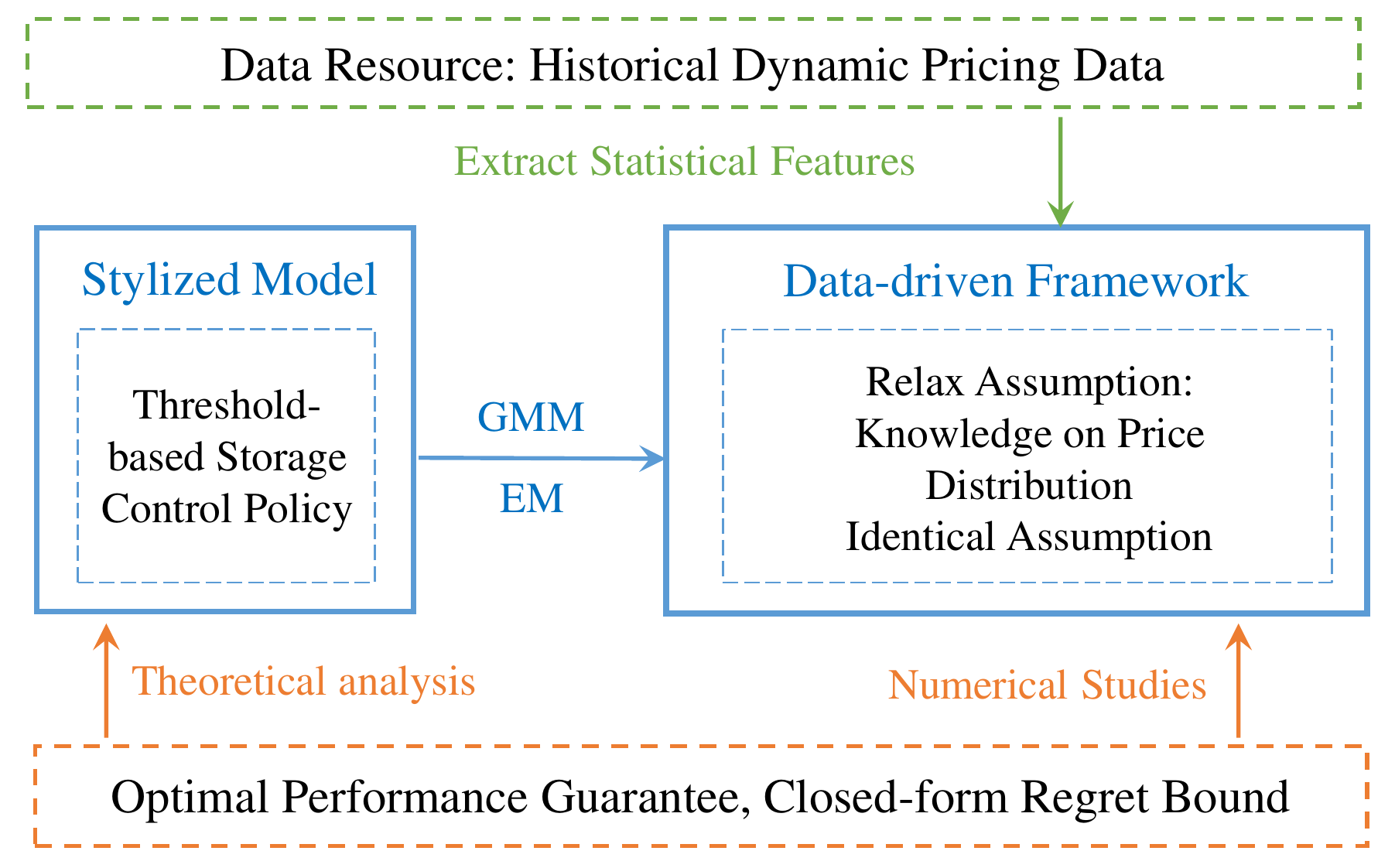}
\caption{Our Data-driven Storage Control Framework}\vspace{-0.2cm}
\label{flowchart}
\end{figure}

Hence, in this paper, to design the storage control framework for demand side facing dynamic prices, we jointly utilize the storage system as well as the intelligence in smart meters. Based on a stylized model, we first identify the optimal threshold-based control policy. With the help of the Gaussian Mixture Model (GMM) \cite{murphy2012machine}, we design a data-driven storage control framework, which is able to further relax the assumptions in the stylized model, e.g., knowledge on the type of distribution, identical assumptions, etc. We plot the flowchart in designing the control framework in Fig. \ref{flowchart}.

\subsection{Literature Review}
Storage system can provide the urgently needed flexibility in the power system. Hence, it has been well investigated to utilize storage system to provide various services in the electricity sector, including smoothing the dispatched wind power generation \cite{5071241Teleke}, maintaining system reliability \cite{op6102369}, smart load scheduling for a single end user \cite{6477197Ven} or a networked community of cooperative users \cite{HatamiDynamic}, grid level operation for minimum cost \cite{Hu2017}, etc.

However, the investigation on storage control policy for dynamic pricing only emerges recently. Jin \textit{et al.} propose a heuristic algorithm using Mixed Integer Linear Programming to optimize the electric vehicle charging schedules in \cite{6461500}. Oudalov \textit{et al.} focus on conducting peak load shaving and introduce a sizing methods as well as an optimal operational scheme in \cite{4538388oudalov}. Chau \textit{et al.} assume the knowledge of future demand and the bounds of prices, and illustrate a threshold-based cost minimizing online algorithm with worst-case performance guarantee in \cite{chau2016cost}. To deal with limited information and uncertainty, Qin \textit{et al.} introduce an online modified greedy algorithm for storage control in \cite{qin2015online}. Vojvodic \textit{et al.} design a simple forward thresholds algorithm to manage the operation in real-time energy market, where they decompose coupling stages using integer programming and heuristic search in \cite{vojvodic2016forward}. Harsha \textit{et al.} propose an optimal storage management policy based on the stochastic dynamic programming and show the policy has a dual threshold structure under mild assumptions in \cite{6880417Harsha}. Different from the literature, we seek to design a \textit{data-driven} storage control framework for dynamic pricing.

Another closely related research line applies data-driven methods to the electricity sector (see \cite{ZHOU2016215} for a systematic survey). Just to name a few, Samuelson \textit{et al.} propose a robust storage control framework to manage the wind power fluctuations in \cite{8062463Samuelson}. Thanos \textit{et al.} design a data-drive simulation framework for automated control in microgrid in \cite{7210221Thanos}. In contrast to most related works, when designing the data driven framework, we provide valuable theoretical insights into the stylized model, which help us understand the value of information in storage control.


\begin{figure}[t]
\centering
\includegraphics[width=2.8in]{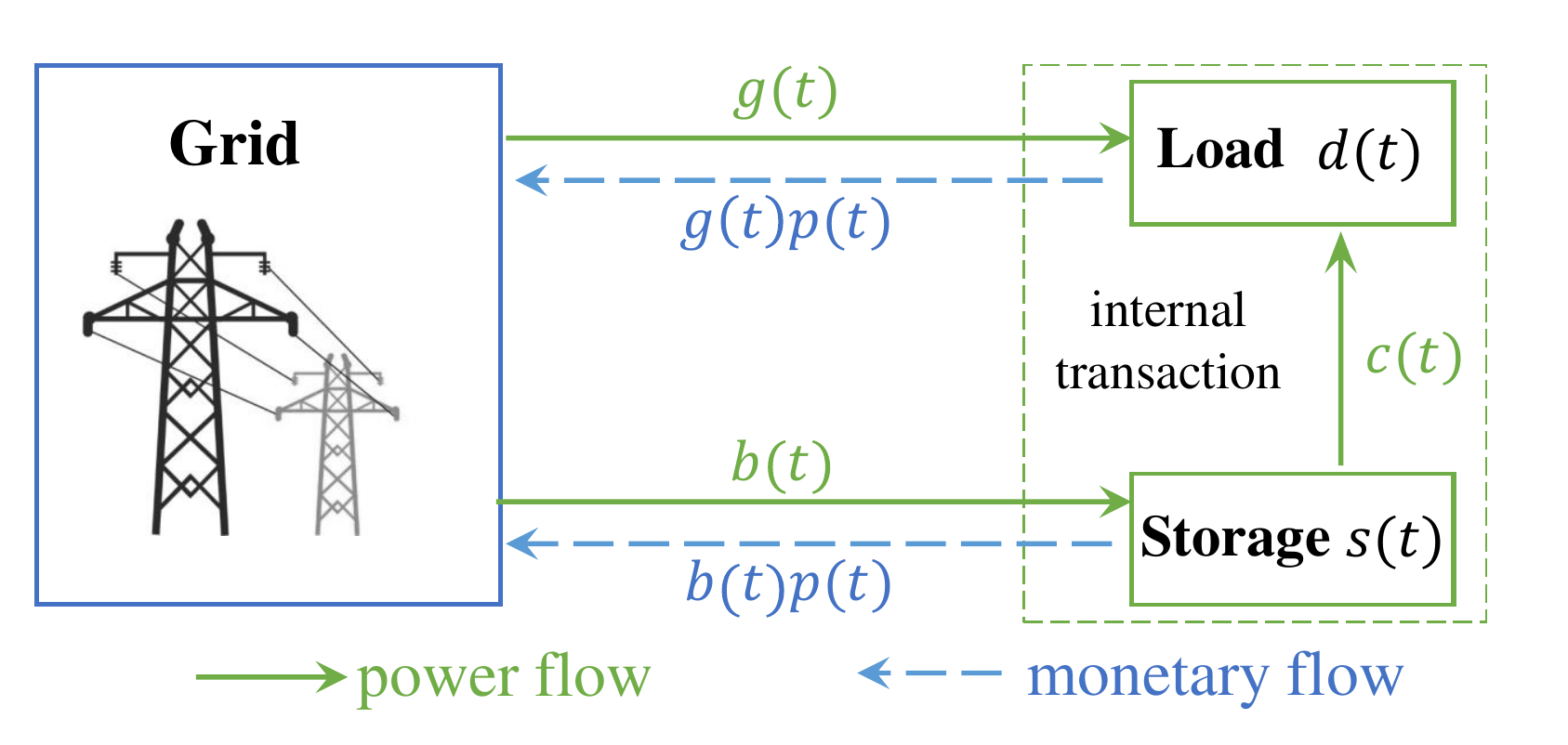}
\DeclareGraphicsExtensions.
\caption{Demonstration of system model.}\vspace{-0.3cm}
\label{fig 2}
\end{figure}

\subsection{Our Contributions}
This paper significantly extends our previous work \cite{our} in both theoretical and practical perspectives. We provide the theoretical basis for the optimal \textit{online} threshold based storage control policy and derive its regret bound, comparing with the \textit{offline} optimal policy. From the practical perspectives, we take a data-driven approach and employ GMM to accurately characterize the price distribution, and use extensive numerical studies to assess the performance of our proposed framework. In summary, we highlight our principal contributions as follows:
\begin{itemize}
    \item \textit{Optimal Online Threshold Policy}: Based on the one-shot load decomposition technique, we propose a simple yet effective online threshold policy to minimize the consumers' expected electricity bills. We prove its optimality and derive its regret bound.
    \item \textit{Data-driven Framework with} GMM: We adopt GMM to relax the assumption of knowing exact price distribution in deriving the threshold policy. With historical price data, we illustrate how to customize the Expectation-Maximization (EM) algorithm \cite{dempster1977maximum} for the parameter estimation in GMM, yielding our data-driven storage control framework.
    \item \textit{Bridge theory and practice}: Our data-driven framework and its heuristic variants enable us to relax many assumptions on the price distributions, including the knowledge of its exact form and the \textit{i.i.d.} assumption. Such relaxations significantly improve the practical feasibility of our framework.  
\end{itemize}

The rest of our paper is organized as follows. Section \ref{sec:2profor} introduces the system model, and revisits the one-shot load decomposition technique. Then, in Section \ref{sec: 3eta}, we design two optimal online threshold storage control frameworks: ETA and DETA. ETA is designed for the stylized model, while DETA and its variants relax many assumptions in the stylized model. In Section \ref{sec:4numerical}, we use numerical studies to evaluate the performance of the two frameworks with real data. Concluding remarks are given in Section \ref{sec:con}. 

\section{One-shot Load Decomposition}
\label{sec:2profor}
Consider the interaction between consumers and the grid as shown in Fig. \ref{fig 2}. The grid operator sets dynamic price $p(t)$ at each time $t$. Facing such a pricing scheme, the consumer wants to satisfy its demand $d(t)$ in different ways: directly purchases energy $g(t)$ from the grid, saves energy $b(t)$ in the storage system, or uses the energy in the storage system ($c(t)$ out of $s(t)$ in the storage) to meet its demand.

Specifically, to highlight the impact of uncertainties in dynamic price on the control policy design, we assume the consumer's own demand prediction is accurate in the near future. This is reasonable as on the grid level, the impact of uncertainties in renewables is reflected through the volatility in dynamic prices. Even with this assumption, the decision making for each consumer is still quite challenging due to the uncertainty in future prices and the physical constraints (capacity constraints) coupling all storage control decisions.

Inspired by \cite{chau2016cost}, in this section, we first revisit the one-shot load decomposition technique, which allows us to decouple the storage control actions across time. Then, we formally formulate the one-shot load serving problem.

\begin{figure}[t]
\centering
\includegraphics[width=2.5in]{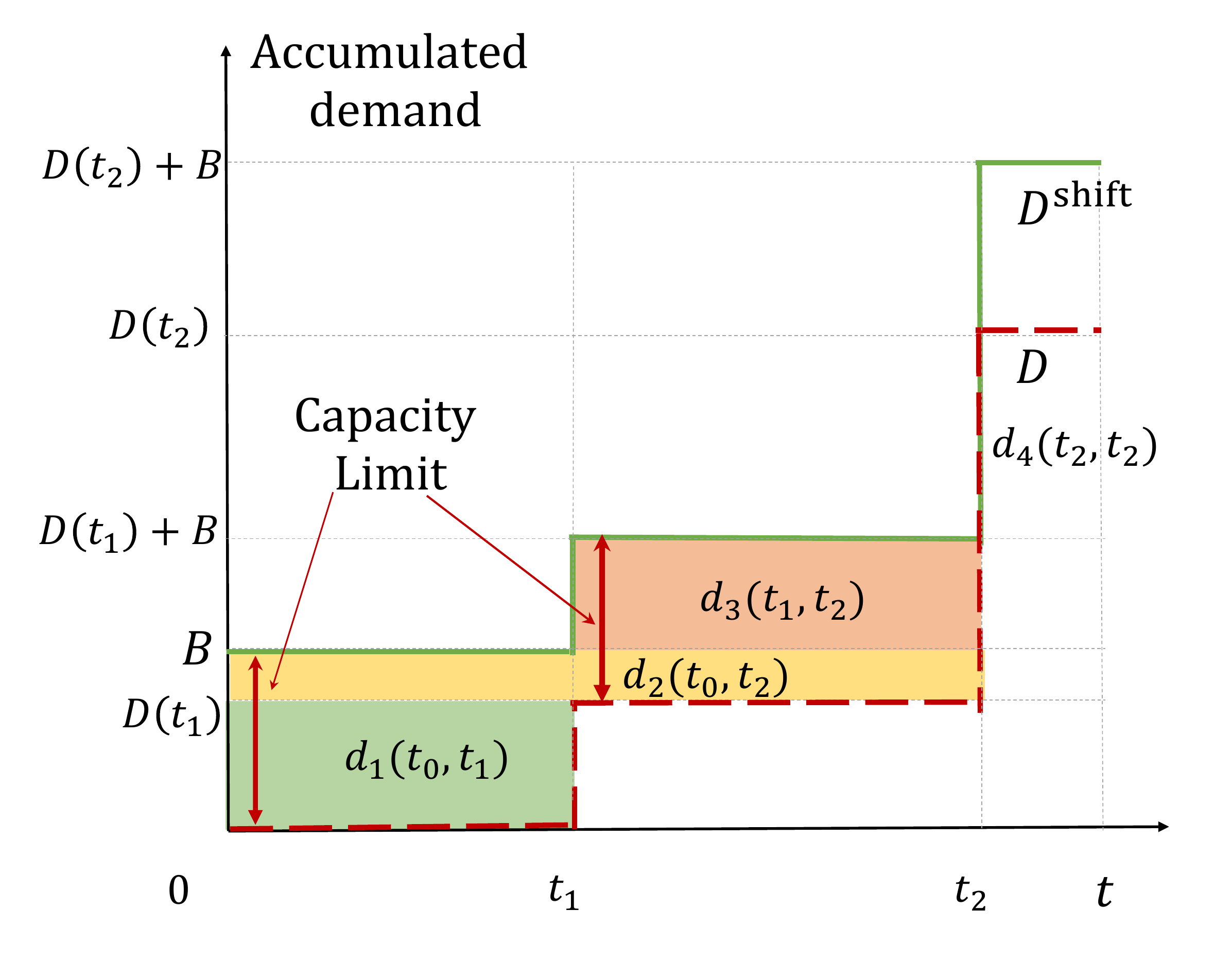}
\DeclareGraphicsExtensions.
\caption{Example for one-shot load decomposition.}\vspace{-0.3cm}
\label{fig 3}
\end{figure}

\subsection{Revisit the One-shot Load Decomposition}
The one-shot load decomposition technique was first proposed by Chau \textit{et al.} in \cite{chau2016cost} to decouple the storage capacity constrained optimization problem into a sequence of one-shot load serving problems.

For example, suppose we need serve an accumulated demand as shown by the red dashed curve in Fig. \ref{fig 3}: serving a load of $D(t_{1})$ at time $t_{1}$, and serving a load of $D(t_{2})-D(t_{1})$ at time $t_{2}$. If there were no storage devices, then we had to purchase the load at the time of serving, tolerating all the price volatility. However, with a storage of capacity $B$, we can reduce such risk.

Back to our example in Fig. \ref{fig 3}, assume
\begin{equation}
B>D(t_{1}):= d_{1}(t_{0},t_{1}),
\end{equation}
then with storage, we can serve $D(t_{1})$
at any time between $[t_{0},t_{1}]$. Therefore, we define $d_{1}(t_{0},t_{1})=D(t_{1})$, which highlights its flexibility in the time span $[t_{0},t_{1}]$. Next, we assume
\begin{equation}
B<D(t_{2})-D(t_{1}),
\end{equation}
then the demand to be met at $t_{2}$ need be decomposed into three different kinds of demand: $d_{2}(t_{0},t_{2})$, $d_{3}(t_{1},t_{2})$, and $d_{4}(t_{2},t_{2})$. Since $B>D(t_{1})$, the storage has certain spare capacity to store energy and serve the load at $t_{2}$ even between $[0,t_{1}]$. This observation leads to the first kind of demand $d_{2}(t_{0},t_{2})$, which is reserved for load at $t_{2}$ between $[0,t_{2}]$ (the union of $[0,t_{1}]$ and $[t_{1},t_{2}]$). The second kind of demand is due to the released capacity after serving load at $t_{1}$, which is only flexible after $t_1$. We denote it by $d_{3}(t_{1},t_{2})$. Note that since $B<D(t_{2})-D(t_{1})$, certain amount of load at time $t_{2}$, $d_{4}(t_{2},t_{2})$, has to be met in real time. This leads to the last type of decomposed demand in the one-shot load decomposition technique. 

\vspace{0.1cm}
\noindent \textbf{Remark}: The last type of decomposed demand naturally decouples the original optimization problem over a \emph{long time} span into a set of \emph{smaller scale} optimization problems. This implies that when we conduct the one-shot load decomposition, we only need look ahead for a couple of hours, the prediction of which can be rather accurate. This also illustrates our assumption of perfect knowledge on the near future demand is realistic.

Next, we formally introduce the construction process for the one-shot load decomposition technique as follows:
\begin{enumerate}
    \item Define the accumulative demand curve $D(t)$.
    \item Define the upward shift accumulative demand curve $D^{\text{shift}}(t)$, which is obtained by shift $D(t)$ by $B$.
    \item Obtain one-shot demand $d_{i}(t_{s}^{i},t_{e}^{i})$ through sandwiched rectangle $(t_{s}^{i}-t_{e}^{i})d_{i}(t_{s}^{i},t_{e}^{i})$ between $D(t)$ and $D^{\text{shift}}(t)$.
\end{enumerate}

\subsection{One-shot Load Serving Problem}
With the one-shot load decomposition, we can focus on solving the one-shot load serving problem. More precisely, for each decomposed load, $d_{i}(t_{s}^{i},t_{e}^{i})$, which need be served between $t_{s}^{i}$ and $t_{e}^{i}$, we seek to find the time $t$ with the minimum price $p(t)$ to serve it. In fact, it suffices for us to understand the stylized one-shot load serving problem, where consumer need satisfy its \textit{one-unit} demand between $1$ and $T$.

Mathematically, if we were able to foresee the future prices, we could directly select:
\begin{equation}
    t^{*}=\mathop{\arg\min}\nolimits_{t\in[1,T]}p(t).
\end{equation}

However, without the perfect knowledge of the future price, we need design an online algorithm to solve the one-shot load serving problem, based on which we can construct the control policy for the general load serving problem.

\section{Optimal Control Framework}
\label{sec: 3eta}

To simplify our subsequent analysis with more insights, we make the following assumption:

\vspace{0.1cm}
\noindent\emph{Assumption}: Dynamic price $p(t)$'s are \textit{i.i.d} random variables.

\vspace{0.1cm}

\noindent With this assumption, we design the control framework for the following two cases:

\begin{itemize}
    \item \textit{Exact Distribution}: Knowing the exact distribution of the random price $p(t)$, we seek to design the optimal online control policy.
    \item \textit{Data-driven Framework}: Based on the optimal control policy, we design the data-driven control framework to relax our assumptions on the price distribution.
\end{itemize}

\subsection{Threshold Control Policy with Exact Distribution}

For the one-shot load serving problem between $[1,T]$, at each time $t$, we only have two choices: to purchase the unit demand or not. The two choices correspond to different expected costs: $p(t)$ for purchasing and $\mathds{E}[w_{t+1}]$ for not purchasing, where $\mathds{E}[w_{t+1}]$ denotes the expected cost for the one-shot unit load serving between $[t+1,T]$. Note that, to calculate $\mathds{E}[w_{t+1}]$, we assume the full knowledge of $p(t)$'s distribution.

To characterize this binary choice, we seek to design the time varying threshold $\theta(t)$ to balance the expected costs between two actions. Hence, the optimal threshold would require
\begin{equation}
\theta(t)=\mathds{E}[w_{t+1}].
\end{equation}
If $p(t) \leq \theta(t)$, we choose to purchase the unit load at time $t$. Otherwise, we defer this action to later time slots.

These thresholds can be obtained in a recursive manner:
\begin{equation}
\theta_{t-1}=\int_{0}^{\theta_{t}}p_{t}f(p)dp+\int_{\theta_{t}}^{\infty}\theta_{t}f(p)dp,
\label{star}
\end{equation}
where $f(p)$ denotes the probability density function of dynamic price $p(t)$, $\forall t$. We have $\theta_{T-1}=\mathds{E}[w_{T}]=\mathds{E}[p(t)]$ as the boundary condition. We term this simple threshold control policy as \textit{expected threshold algorithm} (ETA).

\subsection{Optimality of ETA}
The optimality of ETA comes from the fact that the decision making at each time slot is a binary choice. With this fact, we can prove the following theorem:

\vspace{0.1cm}
\noindent \textbf{Theorem 1}: ETA is the optimal storage control policy for the \emph{one-shot load serving problem}.
\vspace{0.1cm}

\noindent \textbf{Proof}: This theorem can be proved by backward induction. Note that at $\tau=T-1$, $\theta_{T-1} = \mathds{E}[w_{T}]=\mathds{E}[p(T)]$ is the optimal choice. This constructs the induction basis. For the induction part, it suffices to identify that $\theta_{t}$ is the solution to the first order optimality condition of (\ref{star}). \hfill$\blacksquare$


However, it is not obvious why the optimal control policy for the one-shot load serving problem can be used to construct the optimal policy for the original problem. This is because the one-shot load decomposition may change the feasible region of the original problem. Fortunately, we can prove that the one-shot load decomposition maintains the feasible region. Formally, we have the following Lemma.

\vspace{0.1cm}
\noindent \textbf{Lemma 1}: Denote the solution space of general load serving problem by $\mathcal{S}$. Assume this problem can be decomposed into a sequence of $k$ one-shot load serving problems and denote the corresponding solution spaces by $\mathcal{S}_{1},...,\mathcal{S}_{K}$, respectively. It holds:
\begin{equation}
    \mathcal{S} = \bigcup \nolimits_{i=1}^{k}\mathcal{S}_{i}.
\end{equation}

\noindent \textbf{Proof}: Since the decomposition satisfies all the constraints in the original problem, it immediately follows that
\begin{equation}
     \bigcup \nolimits_{i=1}^{k}\mathcal{S}_{i} \subseteq \mathcal{S}.
     \label{ss}
\end{equation}
The difficulty is to show any feasible solution in $\mathcal{S}$ is also in $\bigcup_{i=1}^{k}\mathcal{S}_{i}$. We prove by contradiction. Suppose $\bigcup_{i=1}^{k}\mathcal{S}_{i} \subsetneqq \mathcal{S}$, then there exists a sequence of storage control actions $s=\{\{b(t)\}^{T}_{t=1},\{c(t)\}_{t=1}^{T}\}$, which satisfies the following condition:
\begin{equation}
      s \in \mathcal{S} - \bigcup \nolimits_{i=1}^{k}\mathcal{S}_{i}.
\end{equation}

The fact that $s \in \mathcal{S}$ implies $\sum_{i=1}^{t}b(i)-c(i)\leq D(t)-D(t-1),\forall 0 \leq t \leq T$, while $S \notin \bigcup_{i=1}^{k}\mathcal{S}_{i}$ implies there exists a time $t_{0}$, such that
\begin{equation}
      \sum \nolimits_{i=1}^{t_{0}}b(i)-c(i)>D^{\text{shift}}(t_{0})-D^{\text{shift}}(t_{0}-1).
\end{equation}
However, due to the definition of $D^{\text{shift}}(t)$, we know that
\begin{equation}
      D^{\text{shift}}(t_{0})-D^{\text{shift}}(t_{0}-1)=D(t_{0})-D(t_{0}-1),
\end{equation}
which establishes the contradiction. Hence,
\begin{equation}
     \bigcup \nolimits_{i=1}^{k}\mathcal{S}_{i} \supseteq \mathcal{S}.
\end{equation}

Together with Eq. (\ref{ss}), we complete the proof. \hfill$\blacksquare$
\vspace{0.1cm}

With Lemma 1, the optimality of ETA for the one-shot load serving problem implies its optimality for the general load serving problem. Formally, we state this conclusion as a proposition.

\vspace{0.1cm}
\noindent \textbf{Proposition 1}: ETA is the optimal storage control policy for the general load serving problem.

\subsection{Regret Bound of ETA}
One important metric to evaluate the performance of any online policy is its regret, which measures the performance difference between the online policy and the offline optimal. In this section, since the regret analysis for the general load serving problem depends on the structure of specific problem, we only provide the theoretical regret analysis for the one-shot load serving problem, and conduct numerical studies to highlight ETA's performance for general load serving.

Mathematically, we define the regret of the one-shot load serving problem with shot length $T$ by:
\begin{equation}
\begin{aligned}
 R(T) = \mathds{E}[w]^{\text{ETA}}-\mathds{E}[w]^{\text{Offline}},
\end{aligned}
\end{equation}
where $\mathds{E}[w]^{\text{ETA}}$ and $\mathds{E}[w]^{\text{Offline}}$ represent the expected costs obtained by ETA and offline optimal, respectively. To better characterize the regret $R(T)$, we define the following two parameters:
\begin{equation}
\begin{aligned}
\alpha=\frac{\mathds{E}[p_{2}^{\text{min}}]}{2\mathds{E}[p]},
\end{aligned}
\label{al}
\end{equation}
\begin{equation}
\begin{aligned}
\beta_{t} = \inf \{f(p)|p \in [0,\theta_{t}]\}, 
\end{aligned}
\label{be}
\end{equation}
where $p_{2}^{\text{min}}=\min\{p(1),p(2)\}$. Thus, we can characterize the upper bound of $R(T)$ as follows: 

\vspace{0.1cm}
\noindent \textbf{Theorem 2}: Assume the dynamic price $p(t)$ is non-negative, then, 
\begin{equation}
\begin{aligned}
R(T) \leq \frac{2}{\sum_{i=2}^{T}\beta_{i}}-T\alpha^{T-1}\mathds{E}[p].
\end{aligned}
\end{equation}

We provide the detailed proof in Appendix A. To highlight $R(T)$'s dependence on the moments of dynamic price, we characterize the bound of $R(T)$ in terms of mean and variance for uniform distribution as an illustration.

Assume the dynamic price $p(t)$ follows $U(a,b)$, where $a>0$. Denote the mean and variance of the uniform distribution by $\mu$ and $\sigma^{2}$. Then, standard mathematical manipulation yields
\begin{equation}
\begin{aligned}
 R(T)\leq \frac{4\sqrt{3}\sigma}{T-1}-T\mu\Bigg (1-\frac{\mu^{2}+\sigma^{2}}{2\sqrt{3}\mu\sigma}\Bigg )^{T-1}.
\end{aligned}
\end{equation}

This bound coincides with many of our intuitions. Firstly, $R(T)$ decreases with shot length $T$, of order $O(T^{-1})$. This shows that $R(T)$ approaches $0$ as $T$ becomes large. Secondly, smaller variance (easier to predict) implies better performance guarantee for ETA.

\subsection{Data-driven Control Framework}
We use GMM to relax the assumption of knowing the exact distribution of dynamic price. Due to Central Limit Theorem and the mixture model in GMM, as long as the proper parameters are selected, GMM can approximate any continuous distribution accurately \cite{murphy2012machine}. 


More precisely, GMM assumes that the data is generated from several Gaussian distributions. Hence, its log-likelihood function is of the following form:
\begin{equation}
\begin{aligned}
L(t,K) = \sum\nolimits_{i=1}^{t}\ln\Bigg(\sum\nolimits_{k=1}^{K}\pi_{k}\mathcal{N}(p_{i}|\mu_{k},\sigma_{k})\Bigg),
\end{aligned}
\end{equation}
where $t$ is the number of samples, $K$ is the number of Gaussian components, $\pi_{k}$ is the weight of each component, and $\mathcal{N}(p_{i}|\mu_{k},\sigma_{k})$ is the probability density function for Gaussian distribution, given mean of $\mu_{k}$, and standard deviation of $\sigma_{k}$.

Due to the complex structure in GMM, one primary difficulty in applying GMM for approximating any distribution is the model selection, \textit{i.e.}, how to find the least number of parameters to best represent the data distribution. The reason for selecting the least number of parameters is not only because of Occam's Razor \cite{blumer1987occam}, but also because too many parameters will increase the risk of over-fitting \cite{hawkins2004problem}.

In this paper, we choose to use the Bayesian Information Criteria (BIC) \cite{neath2012bayesian} to determine the optimal number of components in GMM. For a model with $k$ parameters and $t$ samples, its BIC value is defined as follows:
\begin{equation}
\begin{aligned}
BIC(t,k) = k\ln t - 2\ln L(t,k).
\end{aligned}
\end{equation}

For the model with determined number of components, we employ the Expectation-Maximization (EM) algorithm \cite{dempster1977maximum} for parameter estimation. Then we use BIC to select the optimal number of components, which leads to our Data-driven Expected Threshold Algorithm (DETA). To make the algorithm description neat, we only show how to use EM algorithm to estimate the parameters in GMM in Algorithm 1\footnote{In this algorithm, we simply employ the classical EM algorithm. It is possible to design a customized adaptive EM algorithm (see \cite{fan2003semantic} for an example). However, a detailed discussion is beyond the scope of this paper.}. With the estimated GMM as price distribution, we can directly apply ETA for storage control.

\begin{algorithm}
\caption{DETA (Distribution Estimator)}
\label{alg::deta}
\begin{algorithmic}[1]
\REQUIRE $K^{M}$: maximum number of components in GMM;\\
$t$: current time slot;\\
$p(i),i=1,...,t$: historical price data;
\ENSURE
$G^{*}$ : estimated GMM;
\FOR{$K^{m}=$ $1$ $to$ $K^{M}$}
\STATE GMM($t,K^{m}(t)$) Training for $\pi_{k},\mu_{k},\sigma_{k}$,$1\leq k\leq K^{m}$:
Set the initial value of $\pi_{k}$, $\mu_{k}$ and $\sigma_{k}$,$1\leq k\leq K^{m}$.
\REPEAT
\STATE Compute posterior probability:\\ $\gamma(\zeta_{ik}) = \frac{\pi_{k}\mathcal{N}(p_{i}|\mu_{k},\sigma_{k})}{\sum_{j=1}^{K^{m}}\pi_{j}\mathcal{N}(p_{i}|\mu_{j},\sigma_{j})}$\\
\STATE Define $N_{k}=\sum_{i=1}^{t}\gamma(\zeta_{ik})$
\STATE Update parameters according to $\gamma(\zeta_{ik})$ and $N_{k}$:\\
    $\mu'_{k} = \frac{1}{N_{k}}\sum_{i=1}^{t}\gamma(\zeta_{ik})p_{i}$,\\
    $\sigma'_{k} = \frac{1}{N_{k}}\sum_{i=1}^{t}\gamma(\zeta_{ik})(p_{i}-\mu'_{k})(p_{i}-\mu'_{k})$,\\
    $\pi'_{k} = \frac{N_{k}}{t}$.\\
\UNTIL $\mu_{k}$,$\sigma_{k}$,$\pi_{k}$ and $L(t,K^{m})$ converge
\ENDFOR
\STATE Model Selection : choose the model $G^{*}$ as GMM($t,k^{*}(t)$) with \\$k^{*}(t)= \mathop{\arg\min}\limits_{k\in[1,K^{M}]} BIC(t,k)$;
\end{algorithmic}
\end{algorithm}

\subsection{Heuristic Variants for Performance Improvement}
Our \textit{i.i.d.} assumption on the price distribution is for a neat analysis. In fact, in ETA, we needn't restrict ourselves to \textit{i.i.d.} price distribution. Assume a temporally dependent price distribution $f_{t}(p)$ for $p(t)$ at time $t$. Then, we can straightforwardly generalize the threshold recursive structure:
\begin{equation}
\begin{aligned}
\theta_{t-1} = \mathds{E}[w_{t}]=\int_{0}^{\theta_{t}}pf_{t}(p)dp+\int_{\theta_{t}}^{\infty}\theta_{t}f_{t}(p)dp.
\end{aligned}
\end{equation}

The resulting threshold policy can again guarantee the optimality for one-shot load serving problem, and hence, with Lemma 1, the optimality for general load serving. The difficulty lies in the description for price distributions for all time $t$. This motivates us to design the following two heuristic algorithms, both with easy descriptions for price distributions.



One immediate intuition is that the price traces show strong periodicity. The naive period can be 24 hours \footnote{To achieve better performance, one may conduct more comprehensive hypothesis test to find the ``true" period in real data. However, a detailed discussion is beyond the scope of this paper. More importantly, the ``true" period may not necessarily lead to a better performance.}. To capture such periodicity, we can use $24$ GMMs: one for each hour in the day. We term this approach by DETA$^{P}$, where the superscript $P$ emphasizes the periodicity.

The major problem with \textit{i.i.d.} assumption for our storage control is the identical assumption. Another heuristic is to divide each day into a couple of periods, and assume the price distributions within each period are identical. Hence, we adopt a simple adaptive approach: by observing the training period, we first denote the mean of the observed prices by $\mu$. And for each hour, as long as the mean of training price in this hour exceeds $\mu$, we label this hour as peak period. We label all other hours as off-peak periods. This allows us to train only 2 GMMs: one for peak periods, the other for off-peak periods. We term this approach by DETA$^{I}$, where the superscript $I$ emphasizes that we relax the identical assumption in the stylized model. We describe the procedure of DETA$^{I}$ in detail in Algorithm 2.

\begin{algorithm}[h]
\caption{DETA$^{I}$ (Peak and Off-Peak Period Detecting)}
\label{alg::deta}
\begin{algorithmic}[1]
\REQUIRE $T$: number of days for training;\\
    $p^{i}(t): i=1,...,24,t=1,...,T$: historical price data in day $t$ at hour $i$.
\ENSURE
    $\mathcal{P}$: set of peak periods;\\
    $\mathcal{O}$: set of off-peak periods.
\STATE Initialization: $\mathcal{P}=\phi,\mathcal{O}=\phi$.\\
$\mu=[\sum_{i=1}^{24}\sum_{t=1}^{T}p^{i}(t)]/(24T)$
\FOR{$i=$ $1$ $to$ $24$}
\STATE $\hat{\mu} = \frac{\sum_{t=1}^{T}p^{i}(t)}{T}$
\IF{$\hat{\mu}>\mu$}
\STATE $\mathcal{P} \leftarrow \mathcal{P}\bigcup\{i\}$.
\ENDIF
\ENDFOR
\STATE $\mathcal{O}\leftarrow\{1,...,24\}-\mathcal{P}$
\RETURN $\mathcal{P}$ and $\mathcal{O}$
\end{algorithmic}
\end{algorithm}

\vspace{0.1cm}
\noindent \textbf{Remark}: To achieve a better performance, one may want to select a better threshold, rather than the mean $\mu$, to differentiate peak periods and off-peak periods. For example, in our subsequent analysis, we find that selecting a threshold of $40\%$ percentile achieves a better performance. However, we choose not to dive into this more engineering direction. Instead, we emphasize that a simple implementation already achieves improved performance in many cases.

\begin{figure}[t]
\centering
\includegraphics[width=2.5in]{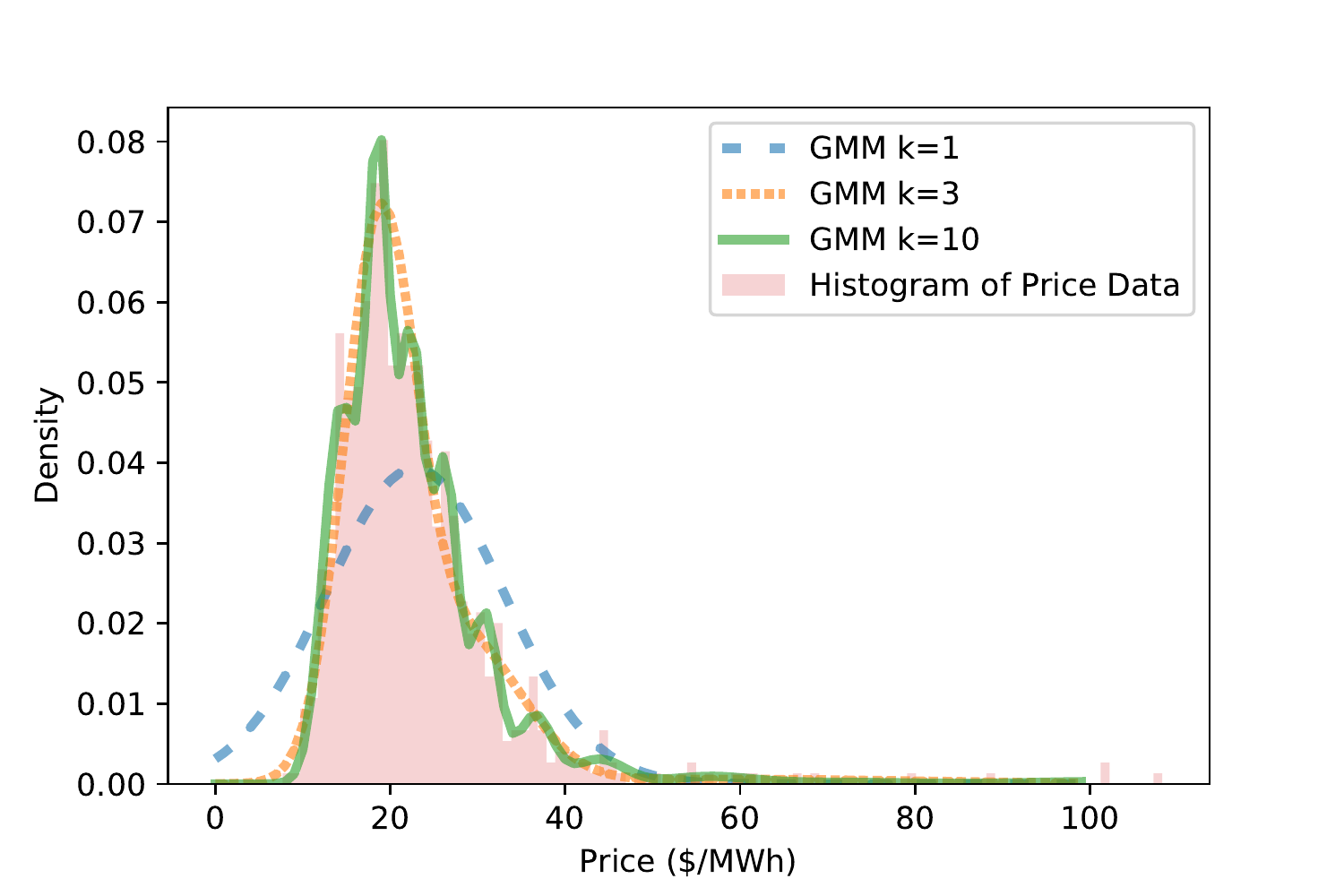}\vspace{-0.2cm}
\DeclareGraphicsExtensions.
\caption{Real price data and its distribution fittings.}\vspace{-0.4cm}
\label{pricedata}
\end{figure}

\section{Numerical Studies}
\label{sec:4numerical}
\begin{figure}[t]
\centering
\includegraphics[width=2.5in]{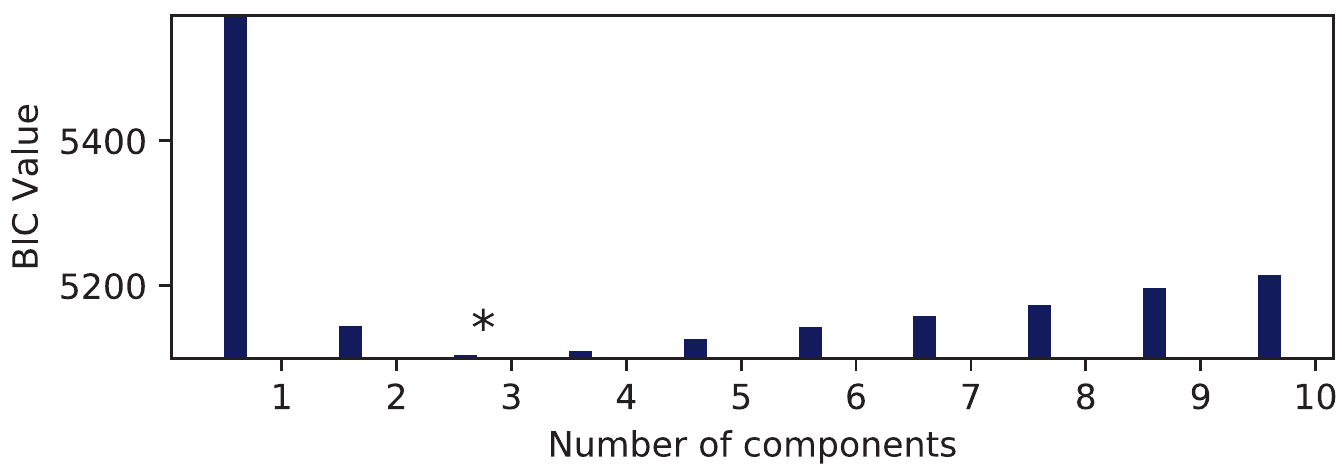}\vspace{-0.2cm}
\DeclareGraphicsExtensions.
\caption{BIC value corresponding to different number of components.}\vspace{-0.2cm}
\label{bic}
\end{figure}

We use the hourly real-time price data to characterize the stochastic nature in dynamic price. The data is collected from AEP \cite{pricedata} during August, 2019. Figure \ref{pricedata} plots the histogram of the prices and its GMM approximation. Figure \ref{bic} suggests that $k=3$ achieves the minimal BIC value (marked by *). Hence, we select a GMM with 3 components for the distribution fitting. To align with the $i.i.d.$ assumption of the price distribution in our stylized model, we first use synthetic data (generated by the fitted GMM) to evaluate ETA's performance. Then, we use real price data to evaluate the performance of DETA and its heuristic variants.As for load dataset, we randomly sample a period from the AEP users' load data, during 2019 \cite{demanddata}.




\subsection{Evaluate ETA for One-shot Load Serving}
\begin{figure}[t]
\centering
\includegraphics[width=2.5in]{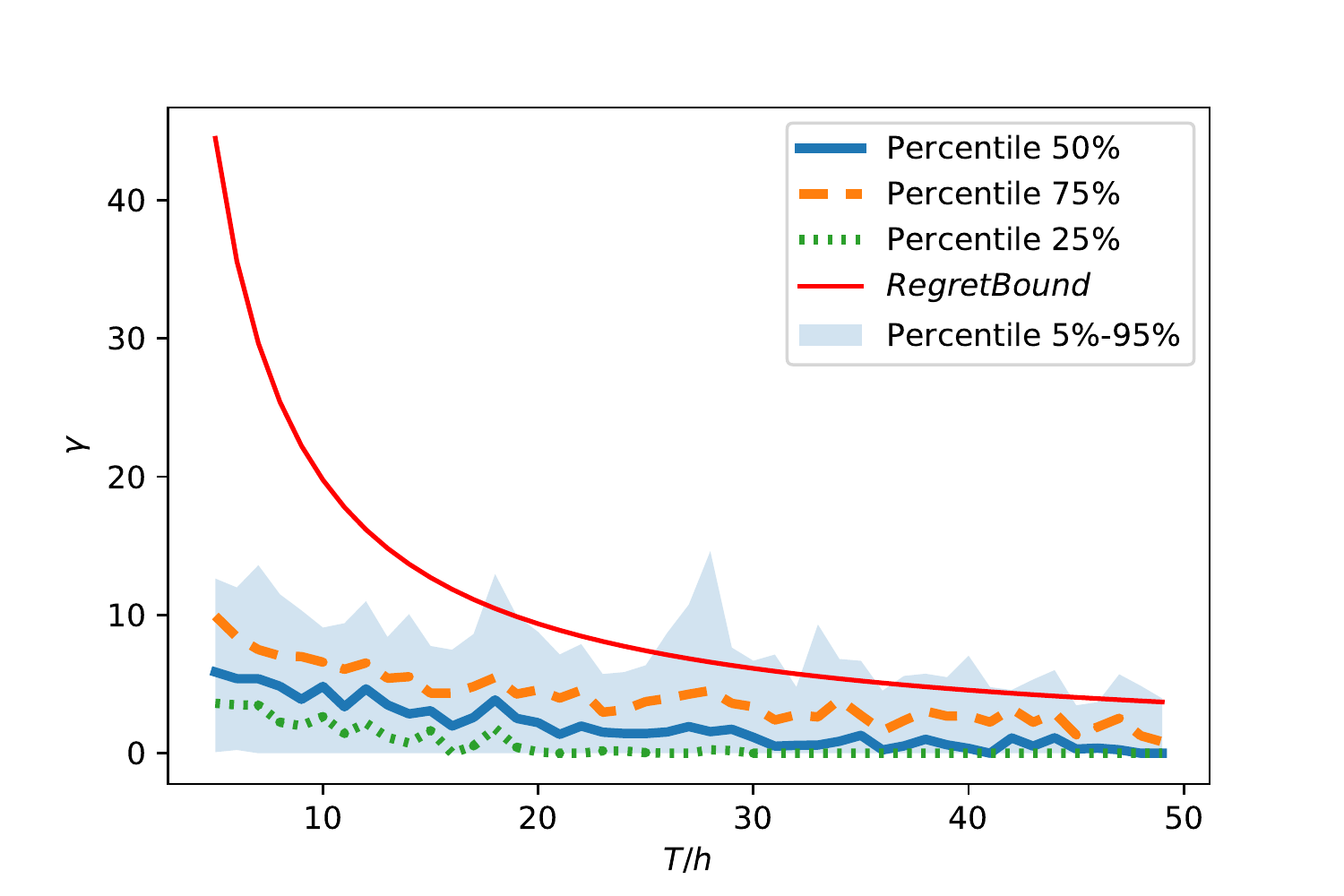}\vspace{-0.3cm}
\DeclareGraphicsExtensions.
\caption{ETA performance for one-shot load serving.}\vspace{-0.3cm}
\label{shot_1}
\end{figure}

We first evaluate the performance of ETA for the one-shot load serving problem, and set the storage capacity to be $10\%$ of the peak hourly energy consumption\footnote{A detailed discussion on optimal storage sizing is provided in Appendix B, which heavily depends on the amortized cost of storage systems.}. To better visualize the comparison between our ETA performance and offline optimal, we define regret ratio $\gamma$ as follows:
\begin{equation}
\gamma = \frac{cost(ETA)-OPT}{OPT},
\end{equation}
where $cost(\text{ETA})$ denotes the cost of ETA, and OPT denotes the offline minimal cost. Figure \ref{shot_1} plots the performance of ETA using the synthetic data, which displays diminishing regret ratio, and implies ETA converges to the offline optimal rather fast in the one-shot load serving problem. 

\subsection{Evaluate ETA for General Load Serving}
\begin{figure}[t]
\centering
\includegraphics[width=2.5in]{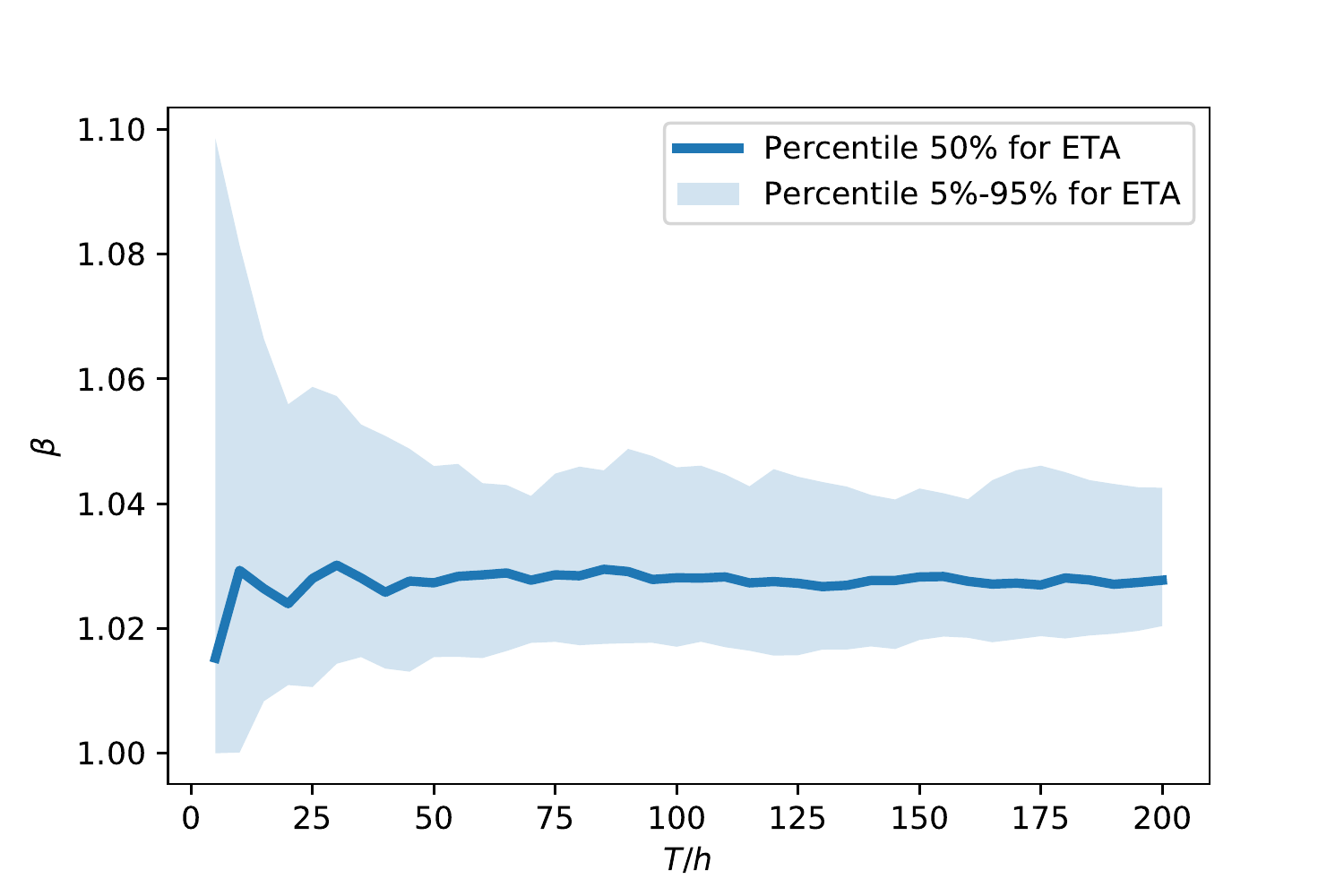}\vspace{-0.2cm}
\DeclareGraphicsExtensions.
\caption{ETA performance for general load serving.}\vspace{-0.2cm}
\label{all_2}
\end{figure}

Next, we evaluate the performance of ETA for general load serving purposes. We define the competitive ratio $\beta$ as follows
\begin{equation}
\beta = \frac{cost(ETA)}{OPT},
\label{competitiveratio}
\end{equation}
where $cost(ETA)$ denotes the \textit{total} cost of ETA in serving load during certain period of time, while OPT denotes the corresponding offline minimal \textit{total} cost. Figure \ref{all_2} plots the competitive ratio for synthetic price data. We observe that the competitive ratio becomes stable as time goes by. Figure \ref{all_2} suggests that the mean of $\beta$ is bounded by $1.04$.




\subsection{Empirical Competitive Ratio}
\begin{figure}[t]
\centering
\subfloat[B = 20\% of peak hourly energy consumption.]{\includegraphics[width=2.7in]{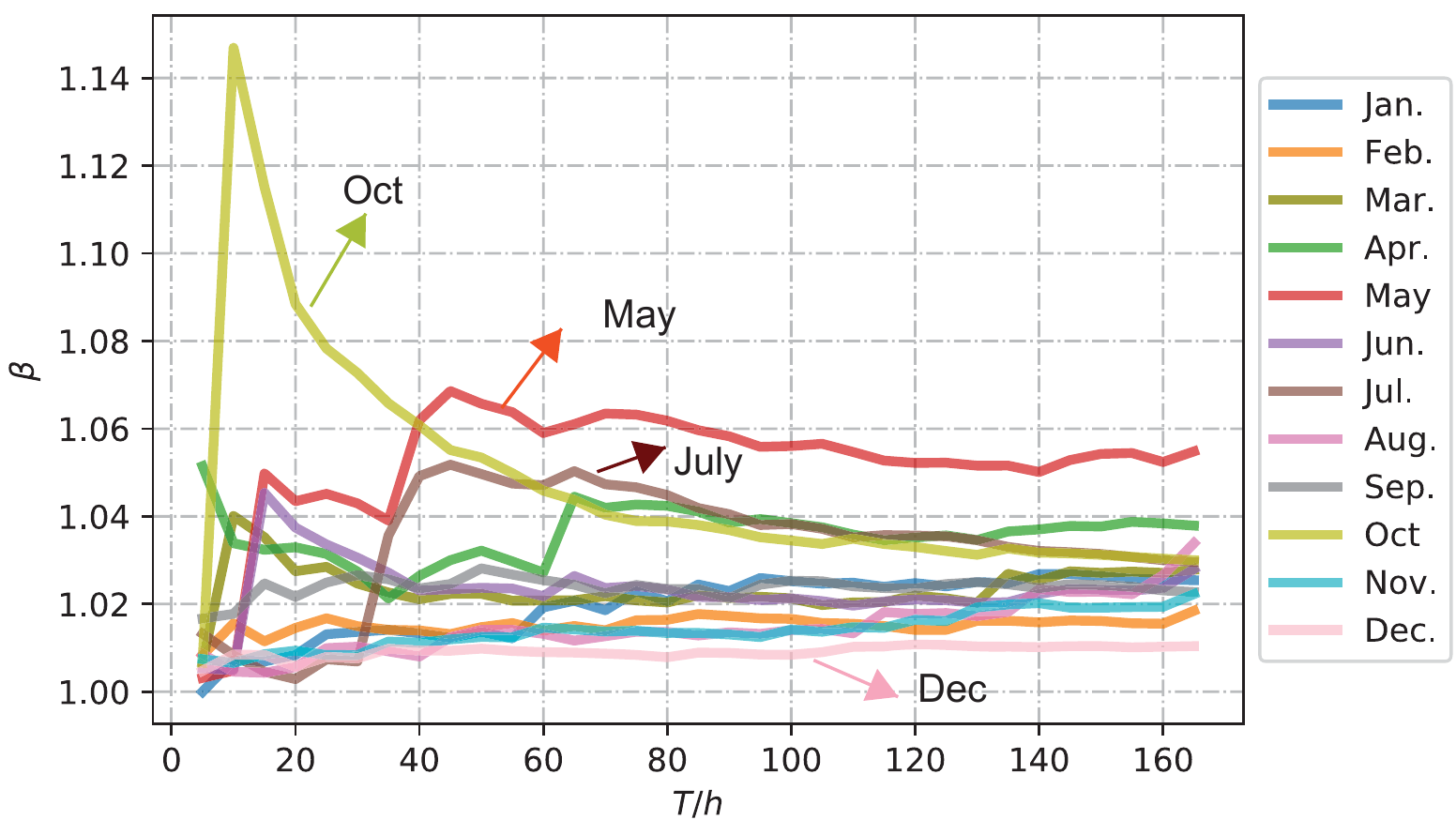}}\vspace{-0.2cm}
\hfil
\subfloat[B = peak hourly energy consumption.]{\includegraphics[width=2.7in]{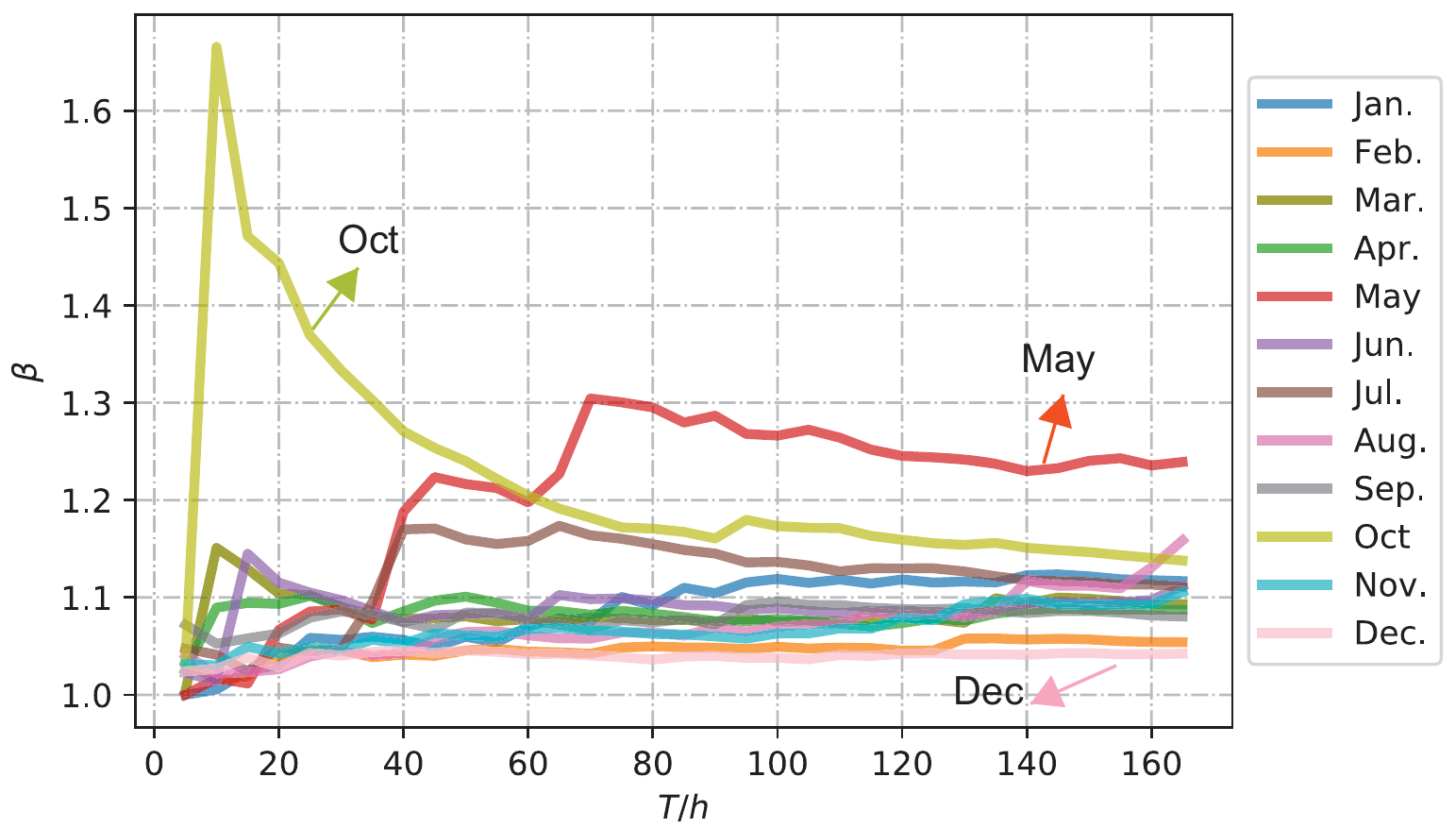}}
\caption{Competitive Ratio of DETA.}\vspace{-0.3cm}
\label{crdeta}
\end{figure}

Next, we use real price data to examine the DETA's competitive ratio in the twelve months. In each month, we use the price data in the first three weeks to train the GMM parameters, and evaluate DETA's performance for the last week. Note that, although this method is not adaptive enough, it is more practical due to its minimal intelligent requirement on the local control devices. As shown in Fig. \ref{crdeta}, the DETA performs remarkably good on all traces. When there is limited flexibility (B is only $20\%$ of the peak hourly energy consumption), the offline optimal cannot save too much. In this case, the average stable competitive ratio is around $1.03$. As there are more flexibility ($B$ is the peak hourly energy consumption), the offline optimal achieves much more savings. In this case, the average stable competitive ratio is around $1.1$.

\begin{figure}[t]
\centering
\subfloat[May]{\includegraphics[width=2.5in]{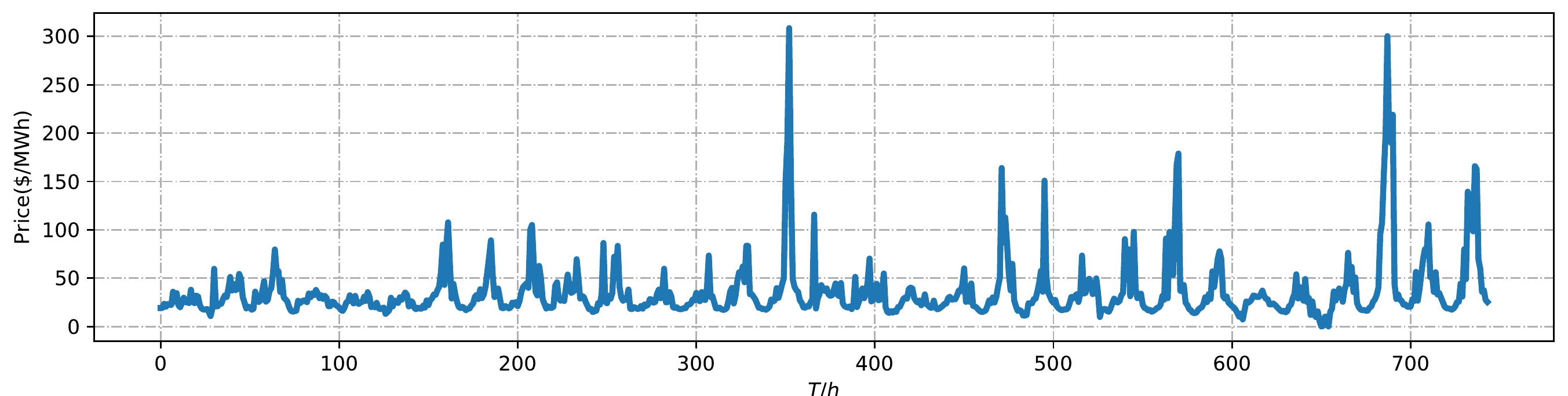}}\vspace{-0.2cm}
\hfil
\subfloat[July]{\includegraphics[width=2.5in]{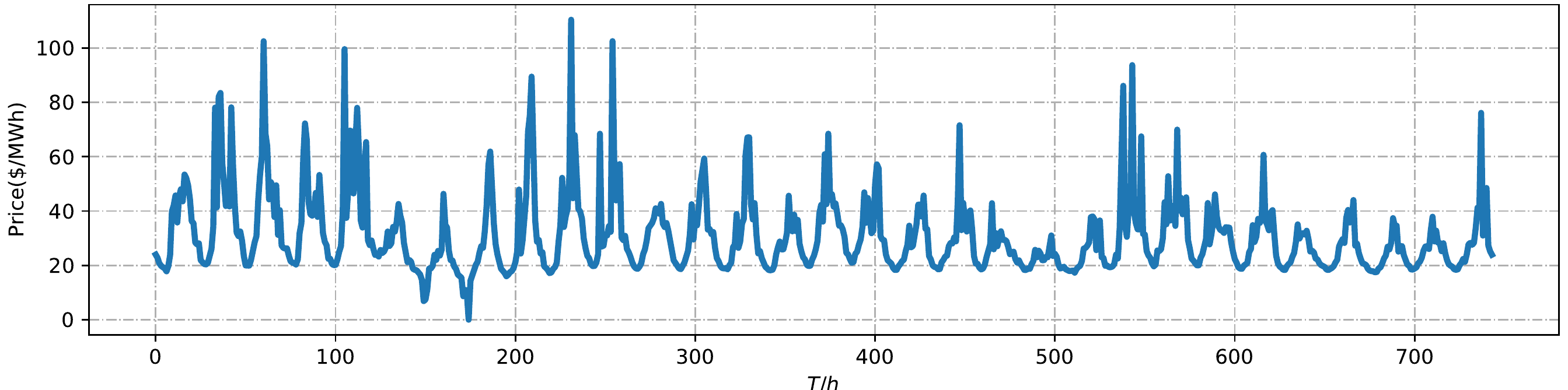}}\vspace{-0.2cm}
\hfil
\subfloat[October]{\includegraphics[width=2.5in]{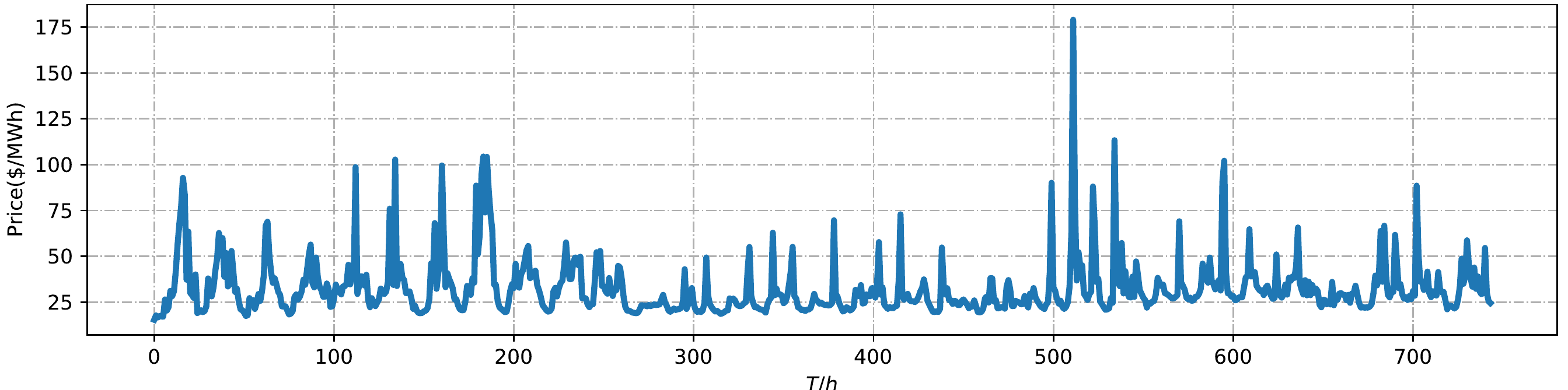}}
\caption{Price Traces in May, July and October.}\vspace{-0.5cm}
\label{price459}
\end{figure}

Although remarkable, in power system control, every $1\%$ saving may be in the order of millions of dollars. Hence, We identify three months with the worst performance in terms of competitive ratio: May, July and October. Figure \ref{price459} further visualizes the price traces for these three months, which can help explain the diverse performance of DETA's heuristic variants.

\begin{figure*}[t]
\centering
\subfloat[May]{\includegraphics[width=1.8in]{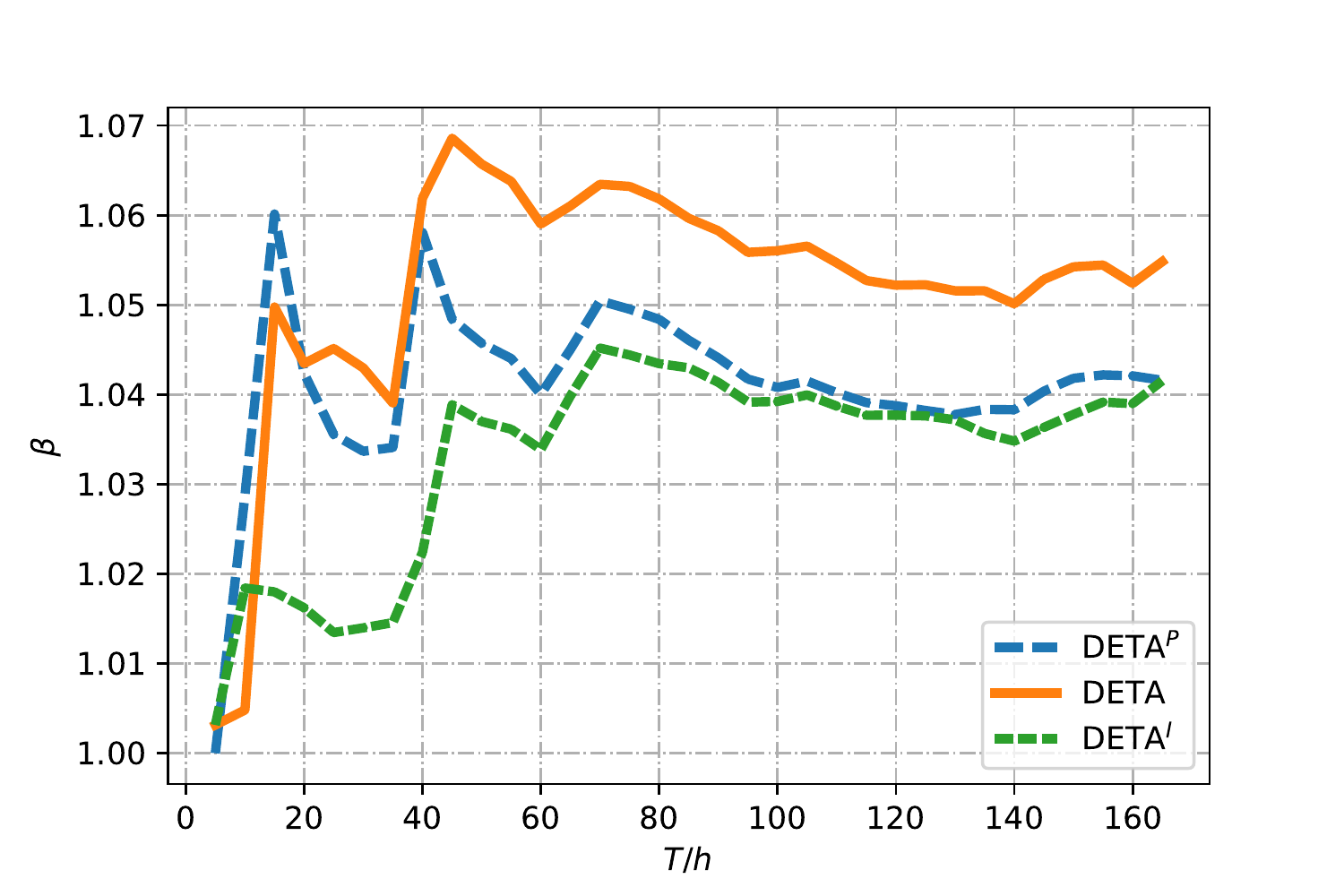}}\vspace{-0.1cm}
\hfil
\subfloat[July]{\includegraphics[width=1.8in]{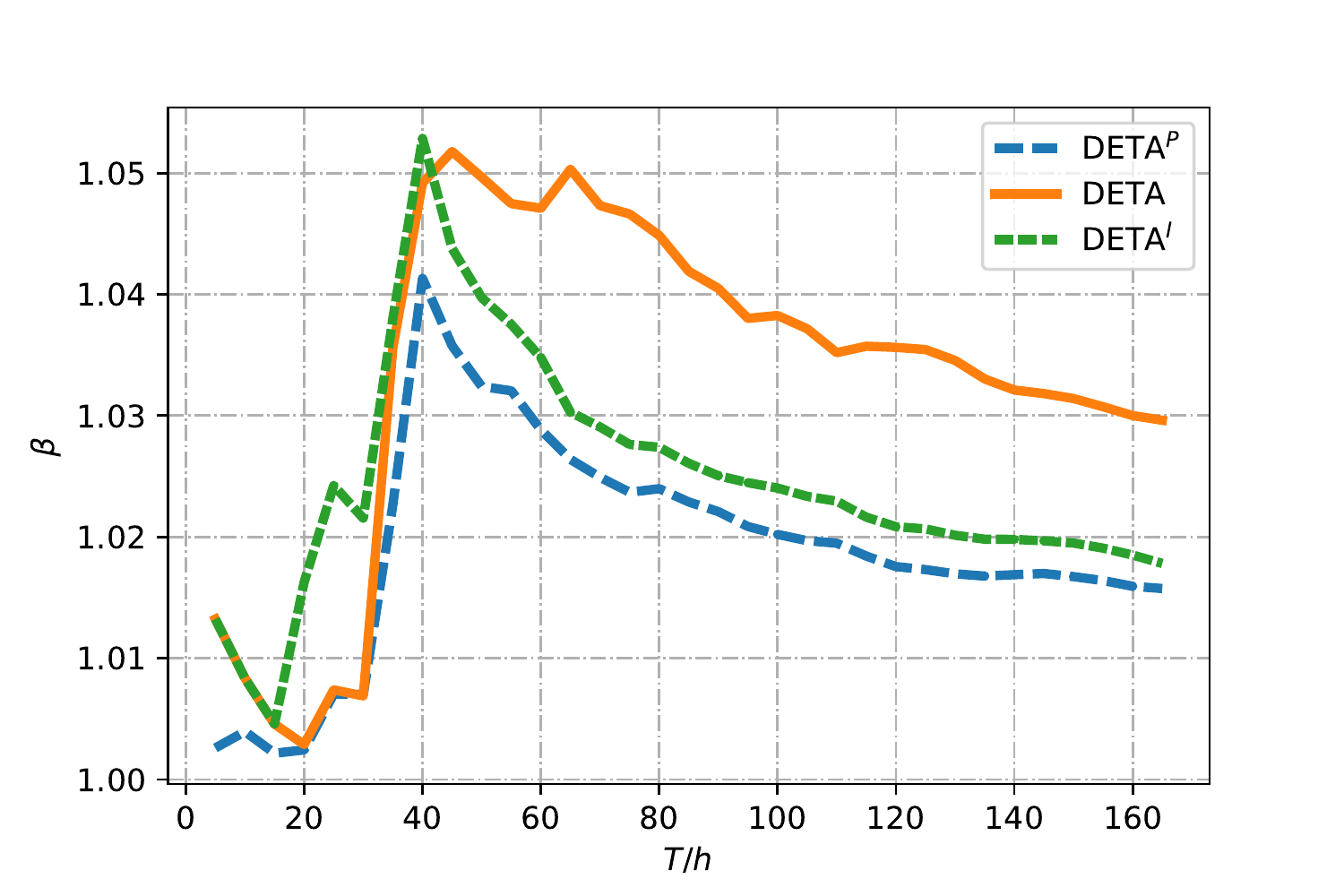}}
\hfil
\subfloat[October]{\includegraphics[width=1.8in]{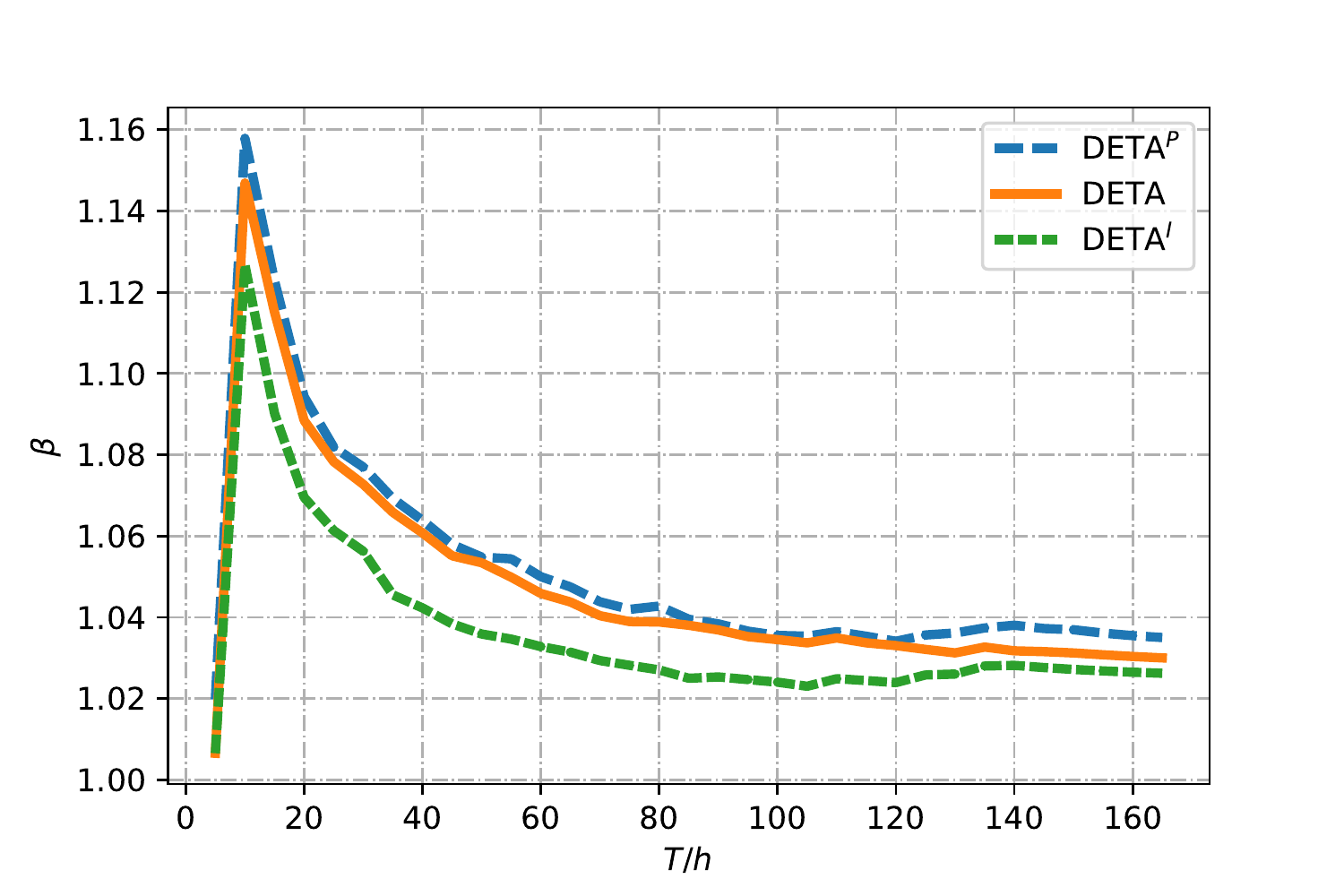}}
\caption{Performance Improvement by Heuristics (B = $20\%$ of peak hourly energy consumption).}\vspace{-0.5cm}
\label{comp1}
\end{figure*}

\begin{figure*}[t]
\centering
\subfloat[May]{\includegraphics[width=1.8in]{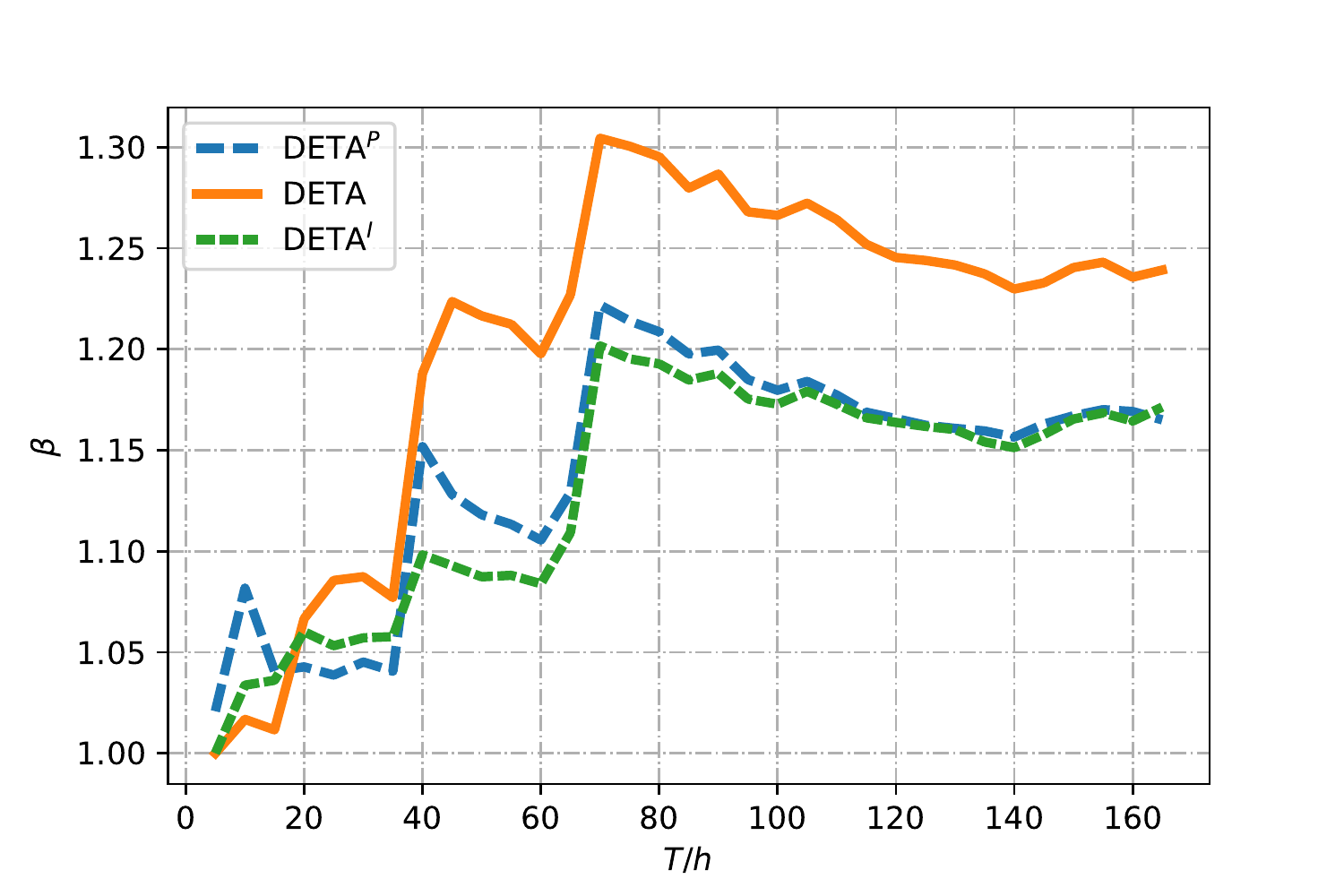}}
\hfil
\subfloat[July]{\includegraphics[width=1.8in]{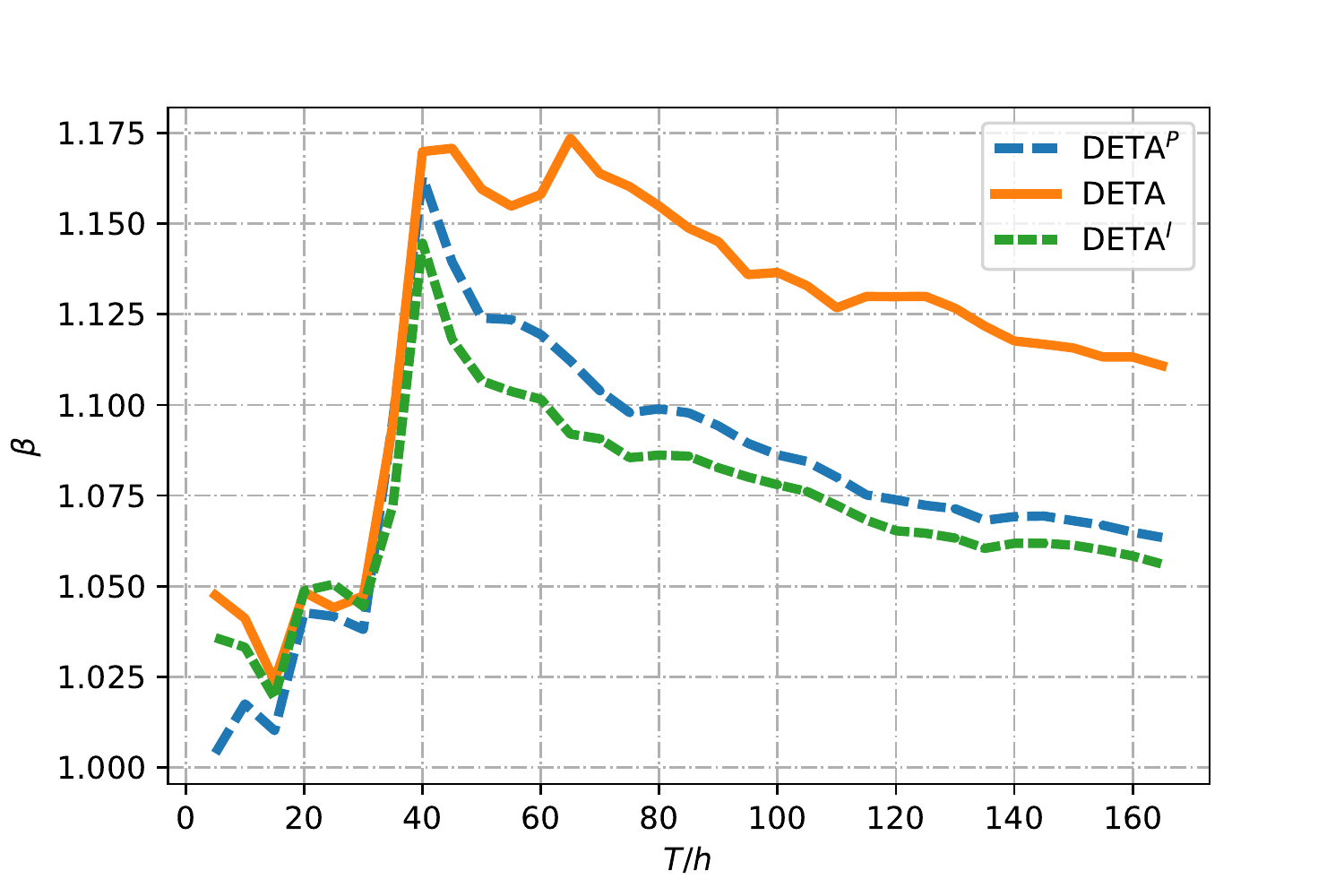}}
\hfil
\subfloat[October]{\includegraphics[width=1.8in]{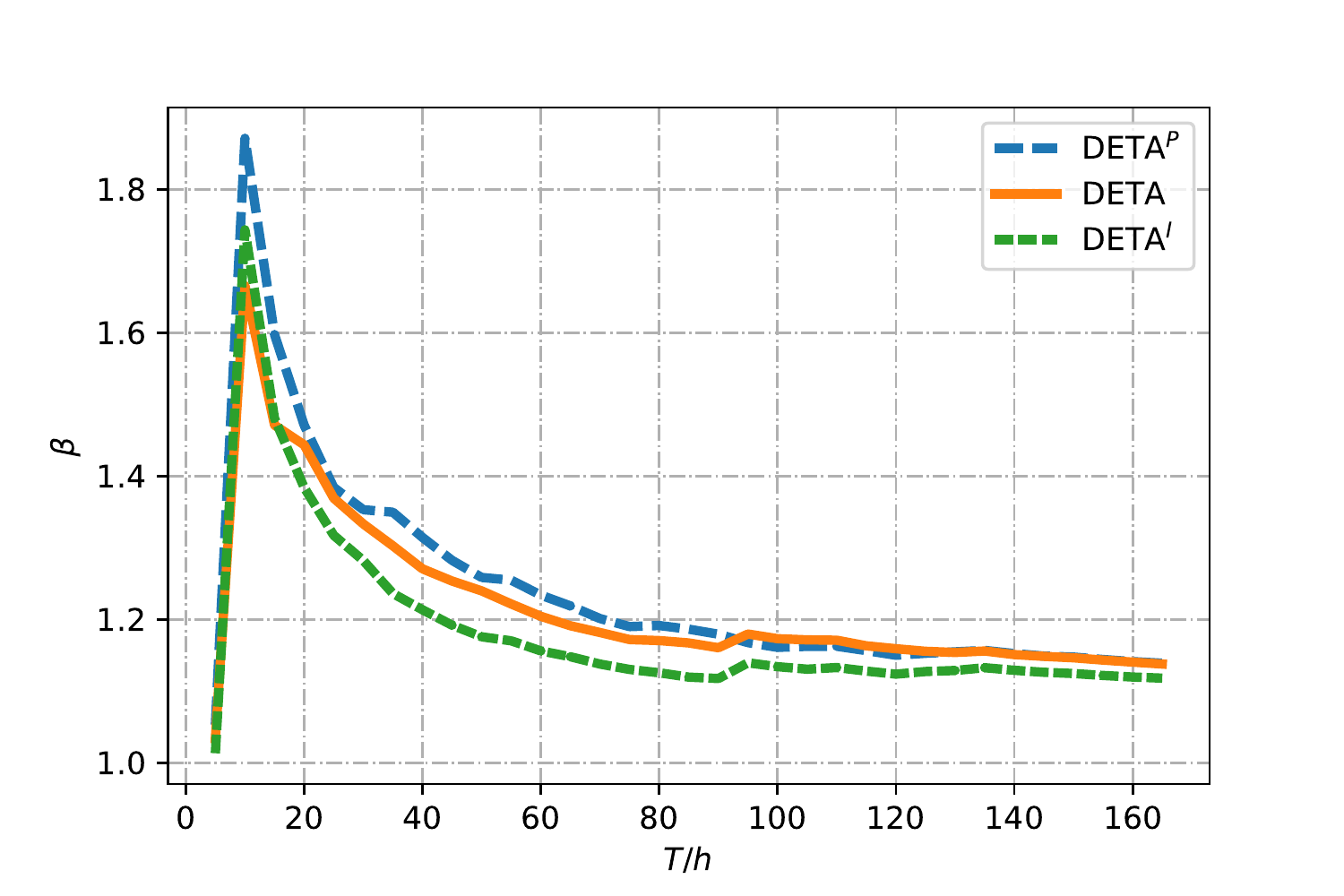}}
\caption{Performance Improvement by Heuristics (B = peak hourly energy consumption).}
\label{comp2}
\end{figure*}

\subsection{Performance Improvement by Heuristics}
We evaluate whether the two heuristic algorithms can actually improve the DETA performance in the three months. We again use the price data in the first $3$ weeks for training, and then test the performance of the three approaches: DETA, DETA$^{P}$, and DETA$^{I}$ (with $40\%$ percentile) for the last week. Figures \ref{comp1} and \ref{comp2} illustrate this comparison, for different storage capacities.  

It is interesting to note that DETA$^{P}$ improves DETA's performance in May and July, but performs poorly in October. This may be because, for example, in October, the price data does not display strong periodicity, comparing with that in the other two months. Also, DETA$^{P}$ requires training $24$ GMMs, which significantly limits the available training data for each GMM.

In contrast, DETA$^{I}$ leads to remarkable improvements in all cases. This implies the potential benefits of improving the current implementation of DETA$^{I}$, by smartly selecting multiple thresholds.

\section{Conclusion}
\label{sec:con}
In this paper, for a stylized model, we design the optimal storage control policy with theoretical guarantee. We generalize the control policy by employing GMM and propose the data-driven framework, which performs remarkably, suggested by numerical studies with real data. We design two promising heuristic algorithms to further enhance the performance of the data driven framework.

This work can be extended in many ways. It is interesting to understand the conditions under which the two heuristic algorithms can achieve the maximal improvement. We also intend to relax the assumption of perfect prediction for the future demand, and study the impact of coupled uncertainties (from both dynamic pricing and the demand) on the optimal control policy design.

\bibliographystyle{ieeetr}
\bibliography{ref}

\section*{Appendix}
\subsection{Proof for Theorem 2}
\noindent \textbf{Proof}
To obtain the upper bound for $R(T)$, we establish the lower bound of $\mathds{E}[w]^{\text{offline}}$ and the upper bound of $\mathds{E}[w]^{\text{ETA}}$, respectively.

Define $p^{\text{min}}_{T} = \min \{p(1),...,p(T)\}$. Then, 
\begin{equation}
  \mathds{E}[w]^{\text{offline}}=\mathds{E}[p_{T}^{\min}]=T\int_{0}^{\infty}p\Bar{F}^{T-1}(p)f(p)dp, 
\end{equation}
where $\Bar{F}(p)$ denotes the tail distribution of $p(t)$. Note that, Cauchy-Schwarz inequality guarantees:
\begin{equation}
\begin{aligned}
    \int_{0}^{\infty}p\Bar{F}^{T}(p)f(p)dp\int_{0}^{\infty}p\Bar{F}^{T-2}(p)f(p)dp \\ \geq \Bigg(\int_{0}^{\infty}p\Bar{F}^{T-1}(p)f(p)dp \Bigg)^{2}.
\end{aligned}
\end{equation}
Hence, we can show that
\begin{equation}
\begin{aligned}
\frac{\mathds{E}[p_{T+1}^{\text{min}}]}{\mathds{E}[p_{T}^{\text{min}}]}&=\frac{T+1}{T}\frac{\int_{0}^{\infty}p\Bar{F}^{T}(p)f(p)dp}{\int_{0}^{\infty}p\Bar{F}^{T-1}(p)f(p)dp}\\
&\geq \frac{T^{2}-1}{T^{2}} \frac{T\int_{0}^{\infty}p\Bar{F}^{T-1}(p)f(p)dp}{(T-1)\int_{0}^{\infty}p\Bar{F}^{T-2}(p)f(p)dp}\\
&=\frac{T^{2}-1}{T^{2}}\frac{\mathds{E}[p_{T}^{\text{min}}]}{\mathds{E}[p_{T-1}^{\text{min}}]}.
\end{aligned}
\end{equation}

Standard mathematical manipulation yields 
\begin{equation}
\begin{aligned}
    &\frac{(T+1)(T-1)}{T^{2}}\frac{\mathds{E}[p_{T}^{\text{min}}]}{\mathds{E}[p_{T-1}^{\text{min}}]}\\
    \geq& \frac{(T+1)\cdot 1}{T \cdot 2} \frac{\mathds{E}[p_{2}^{\text{min}}]}{\mathds{E}[p]}=\frac{T+1}{T}\alpha,
    \end{aligned}
\end{equation}
where $\alpha$ is defined in Eq. (\ref{al}) and $\alpha<\frac{1}{2}$. Thus, we construct the exponentially decreasing lower bound for $\mathds{E}[w]^{\text{offline}}$:
\begin{equation}
    \mathds{E}[w]^{\text{offline}}=\mathds{E}[p_{T}^{\text{min}}]\geq T\alpha^{T-1}\mathds{E}[p].
\end{equation}

Next, we construct the upper bound for $\mathds{E}[w]^{\text{ETA}}=\theta_{T+1}$. Using the recursive structure of threshold, we can show 
\begin{equation}
    \theta_{T}-\theta_{T+1}=\int_{0}^{\theta_{T}}F(p)dp\geq \frac{1}{2}\beta_{T}\theta_{T}^{2},
\end{equation}
where $\beta_{T}$ is defined in Eq. (\ref{be}). To construct the upper bound, it suffices to notice the following inequality holds:
\begin{equation}
    \frac{1}{\theta_{T+1}}-\frac{1}{\theta_{T}}\geq \frac{\theta_{T}-\theta_{T+1}}{\theta_{T}^{2}}\geq\frac{\beta_{T}}{2}.
\end{equation}
And when $T=1$, we have to purchase for the unit demand at any price. Hence, $\theta_{1}=\infty$. This yields
\begin{equation}
    \theta_{T+1}\leq \frac{2}{\sum_{i=2}^{T}\beta_{i}}.
\end{equation}
As long as $\sum_{i=2}^{T}\beta_{i}>0$, then, the desirable upper bound immediately follows. \hfill$\blacksquare$

\subsection{Storage Sizing Problem}
To decide the optimal storage sizing, it suffices to understand the marginal benefit of storage in this problem, which can be answered by the following parametric analysis:
\begin{equation}
    \begin{aligned}
    \text{MinC}(B)&:=\min \quad  \sum\nolimits_{t=1}^{T}\mathds{E}[(g(t)+b(t))p(t)]\\
\mbox{s.t.}\quad
&g(t)+c(t)=\mathds{E}[d(t)],\quad \forall t\in[1,T],\\
&\sum \nolimits_{\tau=1}^{t} (b(\tau)-c(\tau))\leq B,\quad \forall t\in[1,T]. \label{bcons}
    \end{aligned}
\end{equation}

Notice that the function $\text{MinC}(B)$ is continuous, non-increasing and concave \cite{holder2010parametric}. Hence, the derivative of $\text{MinC}(B)$ is also non-increasing. Denote the amortized cost for storage system over time span $T$ by $\pi_{b}$. Then, the optimal storage capacity $B^{*}$ is the solution to the following equation:
\begin{equation}
    \pi_{b}=\text{MinC}'(B^{*}).
\end{equation}

Thus, by selecting the proper amortized cost, the optimal storage capacity can be immediately obtained.

\end{document}